\documentclass{icrc2009}

\usepackage{graphicx}   
\usepackage{caption}    
\usepackage[font=footnotesize]{subfig} 
\usepackage{fixltx2e}
\usepackage{url}

\newcommand{\shorttitle}[1]%
{\markboth{Proceedings of the 31\MakeLowercase{$^{st}$} ICRC, {\L}\'{o}d\'{z} 2009}{#1} }
\newcommand{\etal}{\MakeLowercase{\textit{et al. }}} 

\begin{document}
\title{Spectrum and composition of galactic cosmic rays accelerated in supernova remnants}

\author{\IEEEauthorblockN{Vladimir Ptuskin\IEEEauthorrefmark{1},
              Eun-Suk Seo\IEEEauthorrefmark{2},
                          Vladimir Zirakashvili\IEEEauthorrefmark{1}}

                            \\
\IEEEauthorblockA{\IEEEauthorrefmark{1}IZMIRAN, Troitsk, Moscow region 142190, Russia}
\IEEEauthorblockA{\IEEEauthorrefmark{2}IPST, University of Maryland, College Park, MD 29742 USA}}

\shorttitle{Ptuskin \etal acceleration in supernova remnants}
\maketitle

\begin{abstract}
 The spectra of high-energy protons and nuclei accelerated by supernova remnant shocks
 are calculated taking into account magnetic field amplification and Alfvenic drift for
 different types of SNRs during their evolution. The overall energy spectrum and elemental
 composition of cosmic rays after propagation through the Galaxy are found.
 \end{abstract}


\section{Introduction}
 The supernova remnants (SNRs) are recognized as the principle sources of
Galactic cosmic rays, and the diffusive shock acceleration is
accepted as the acceleration mechanism of cosmic rays by a
supernova blast wave moving through the turbulent interstellar
medium. Accelerated by supernova shocks, the energetic particles
diffuse in the interstellar magnetic fields and fill the entire
Galaxy. Clear evidence for particle acceleration in SNRs is given
by observations of non{}-thermal radio, X{}-ray, and gamma{}-ray
radiation. It is experimentally established from the H.E.S.S. data
that there are cosmic{}-ray particles with energies exceeding
$10^{14}$ eV \ in the shell of the supernova remnant RX
J1713.7{}-3946 [1].

For a time, the theoretical estimates of maximum proton energy
were at a level of  $E_{\mathrm{max}}{\approx}10^{14}-10^{15}$ eV
[2]. The work [3] demonstrated that a very large random magnetic
field $\mathrm{{\delta}B}\gg B_{\mathrm{ism}}$ (where
$B_{\mathrm{ism}}{\approx}5\mathit{{\mu}G}$ is the average
interstellar field upstream of the shock) can be generated by
cosmic{}-ray streaming instability in the precursor of strong
shocks. It results then in efficient confinement of energetic
particles in the shock vicinity, and it may rise the maximum
particle energy by about two orders of magnitude (the exact value
depends on the supernova parameters, see below). Remarkably, data
on synchrotron X{}-ray emission from a number of young SNRs proved
the presence of strong magnetic fields  $150$  to $500$ ${\mu}$G
[4] that can be naturally explained by the effect of cosmic{}-ray
streaming instability.

In the present work we calculate the steady state spectrum of cosmic
rays in the Galaxy using recent results on magnetic field
amplification in SNRs.\\

\section{Calculation of cosmic ray spectrum in the  Galaxy}

 Because of high efficiency of shock acceleration, the spectrum of
cosmic rays should be selfconsistently determined with the account
of shock modification caused by the pressure of accelerated
particles. We study cosmic ray acceleration and the evolution of a
supernova blast wave with the use of our numerical code described
in [5] (see also the paper "Numerical simulations of shock
acceleration in SNRs including magnetic field amplification" by
Zirakashvili and Ptuskin at this Conference for the improved
version of the code). The hydrodynamic equations are solved
together with the diffusion{}-convection transport equation for
the cosmic ray distribution function $f(t,r,p)$, which depends on
time $t$, radial distance from the point of supernova explosion
$r$ (here the spherical symmetry is assumed), and the particle
momentum $p$. The accepted injection efficiency of thermal ions in
the process of shock acceleration  $\eta =0.1u_{\mathrm{sh}}/c$ is
taken in accordance with [6]; here $u_{\mathrm{sh}}(t)$  is the
varying in time shock velocity and $c$ is the speed of light. The
new essential feature of our calculations is the inclusion of the
Alfvenic drift effect for particle transport. The Alfvenic
velocity
 $V_{A}=B/\sqrt{4\pi \rho }$  is not negligible in comparison to the gas
velocity {\textmd{$u$} downstream of the shock if the magnetic
field is significantly amplified as it follows from the
observations of the synchrotron X{}-ray radiation. The positive
(directed outside) gradient of accelerated particles in the
downstream region from the shock generates Alfven waves
propagating in the negative direction, and we set the cosmic ray
advection velocity equal to $w=u-V_{A}/\sqrt{3}$ there. Due to
this effect, the accelerating particles ``feel'' a smaller
compression ratio and acquire softer energy spectrum compared to
the usual assumption $w=u$. We employ results [4] on the analysis
of X-ray radiation from young SNRs and assume that magnetic energy
density $B^{2}/8\mathrm{{\pi}}$ downstream of
 the shock is $3.5$ \% of the ram pressure
 $\mathit{{\rho}u}_{\mathrm{sh}}^{2}$.
It is worth noting that this relation is in a good agreement with
the modelling of cosmic ray streaming instability in young SNRs
[7].

The Bohm diffusion coefficient
$D_{B}=\mathit{vpc}/(3\mathit{ZeB})$ is assumed for the
accelerating particles of charge $\mathit{Ze}$ and velocity $v$.
The maximum particle momentum $p_{\mathrm{max}}$ reached in a
process of diffusive shock acceleration can be roughly estimated
from the condition $D_{B}(p_{\mathrm{max}})\sim
0.1u_{\mathrm{sh}}R_{\mathrm{sh}}$ with $D_{B}$ calculated for the
upstream magnetic field, which is about $5$ times smaller than the
downstream field. This gives an order of magnitude estimate
$p_{\mathrm{max}}c/Z\sim
24(u_{\mathrm{sh}},_{3})^{2}R_{\mathrm{sh}}\sqrt{n}$ TeV, where
$10^{3}u_{\mathrm{sh}},_{3}$ km/s is the shock velocity,
$R_{\mathrm{sh}}$ pc is the shock radius, $n$ cm${}^{-3}$ is the
interstellar gas number density.

 It can be shown, see [8], that the transformation of supernova
explosion energy to cosmic rays becomes efficient from the
beginning of the Sedov stage of the shock evolution (i.e. when the
mass of supernova ejecta becomes equal to the mass of swept-up
gas) and continues later on. As a result, the characteristic knee
arises in the overall spectrum of particles accelerated by the
evolving supernova remnant. The position of knee
$p_{\mathrm{knee}}$  can be estimated from the above equations for
$p_{\mathrm{max}}$ where $u_{\mathrm{sh}}$  and  $R_{\mathrm{sh}}$
are determined at the time when Sedov stage begins. It gives
approximately

\begin{equation}
p_{\mathrm{knee}}c/Z\sim 1.1\cdot {10}^{15}E_{51}n^{1/6}M_{\mathrm{ej}}^{-2/3}
\mathrm{eV}.\ \
\end{equation}

Here $E_{51}$ is the kinetic energy of the supernova explosion in
units of $10^{51}$ erg and $M_{\mathrm{ej}}$ is the mass of
supernova ejecta measured in the solar masses.

If the presupernova had a dense star wind with velocity $u_{w}$
and the mass loss rate $\dot{M}$ before the explosion, the shock
may enter the Sedov stage while propagating through the wind
material with mass density $\rho _{w}=\dot{M}/(4\pi u_{w}r^{2})$.
Eq. (1) should be replaced in this case by the following equation:

\begin{equation}
 p_{\mathrm{knee}}c/Z\sim 8.4\cdot {10}^{15}E_{51}\sqrt{{\dot{M}}_{-5}/u_{w,6}}M_{\mathrm{ej}}^{-1}
\mathrm{eV}.
\ \
\end{equation}

 We fulfilled numerical simulations of cosmic ray acceleration for
$4$ types of supernova remnants.

$1.$ Type Ia SNRs with the following parameters: the kinetic
energy of explosion $E={10}^{51}\mathrm{erg}$, the number density
of the surrounding interstellar gas $n=0.1$ cm${}^{-3}$, the\ \
mass of ejecta $M_{\mathrm{ej}}=1.4M_{\odot }$. Also important for
accurate calculations is the index $k$ which describes the power
law density profile $\rho _{s}\propto r^{-k}$ of the outer part of
the star that freely expands after supernova explosions; $k=7$ for
Type Ia supernova.

$2.$ Type IIP SNRs with parameters $E$=${10}^{51 }\mathrm{erg}$,
$n=0.1 $ ${\mathrm{cm}}^{-3}$, $M_{\mathrm{ej}}=8M_{\odot }$,
$k=12.$

$3.$ Type Ib/c SNRs with $E={10}^{51}$ erg exploding into the low
density bubble with density $n=0.01$ cm${}^{-3}$ produced by the
progenitor star when it was on the main sequence,
$M_{\mathrm{ej}}=2M_{\odot }$, $k=7$.

$4.$ Type IIb SNRs with $E=3\cdot {10}^{51}$ erg,
$n=0.01$cm${}^{-3}$, $M_{\mathrm{ej}}=1M_{\odot }$. Before
entering the rarefied bubble, the blast wave goes through the
dense wind emitted by the progenitor star during its final RSG
(Red Super Giant) stage of evolution. We assume that the mass loss
rate by the wind is $\dot{M}={10}^{-4}M_{\odot }/\mathrm{yr}$ and
the outer wind radius is $5$ pc.

The discussion about properties of SNRs produced by core collapse
supernovae can be found in [9].

Based on the statistics of supernovae within $28$ Mpc of the
Galaxy [10], we accept the following relative rates for the $4$
types of supernovae described above: $0.30, 0.44, 0.22$, and
$0.04$ respectively.

The calculated cosmic ray spectra produced over the lifetime of
each type of supernovae are shown in Fig.1 under the assumption
that only protons are accelerated. Here $Q(p)=4\pi p^{2}F(p)$,
where $F(p)$ is the distribution of all accelerated particles
injected in the interstellar medium over SNR lifetime. (The total
number of accelerated particles is $\int Q( p) dp$). It was
assumed that the acceleration ceased at $t_{c}={10}^{5}$ yr. The
shock velocity at this moment is close to $200$ km/s and the
maximum energy of protons confined in the supernova remnant is
$\sim 10$ TeV. All particles with higher energies accelerated
earlier have left the remnant. The maximum particle energy at late
stages of shock evolution can be much smaller if one takes into
account the possible damping of turbulence in the shock precursor
due to ion-neutral collisions or non-linear wave interactions
[11].

The function $F(p)$ was calculated as the sum of two integrals:
the integral taken at $t_{c}$ over the volume of supernova remnant
$(4\pi )\int _{0}^{R_{\mathrm{sh}}( t_{c}) }f(t_{c},r,p) r^{2}dr$,
and the integral over time of the diffusion f{}lux of accelerated
particles through the boundary of the calculation domain $[4\pi\
r^{2}\int _{0}^{t_{c}}(-D \partial f/\partial r)dt]|_{r_{b}}$. The
source function $Q( p)$ should be multiplied by $\nu
_{\mathrm{sn}}$, where $\nu _{\mathrm{sn}}$ is the supernova rate
per unit volume to obtain the density of cosmic ray sources. Fig.
1 shows that about $1/3$ of supernova explosion kinetic energy $E$
goes to cosmic rays, which is in agreement with the empirical
model of cosmic ray origin.

\begin{figure}
\includegraphics[width=3.3in]{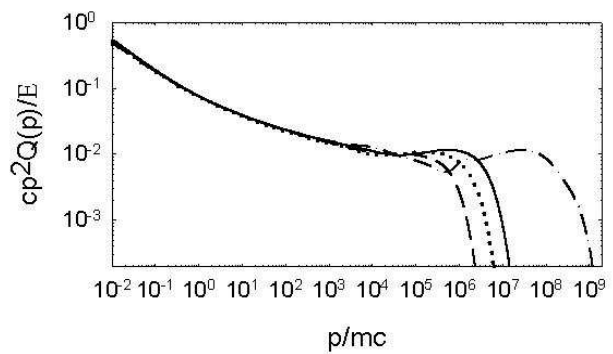}
\caption{Proton source spectra produced by supernovae Type Ia
(solid line), Type IIP (dash line), Type Ib/c (dotted line), and
Type IIb (dash-dot line)}
\end{figure}

\begin{figure*}[!t]
   \centerline{\subfloat[]{\includegraphics[width=2.3in]{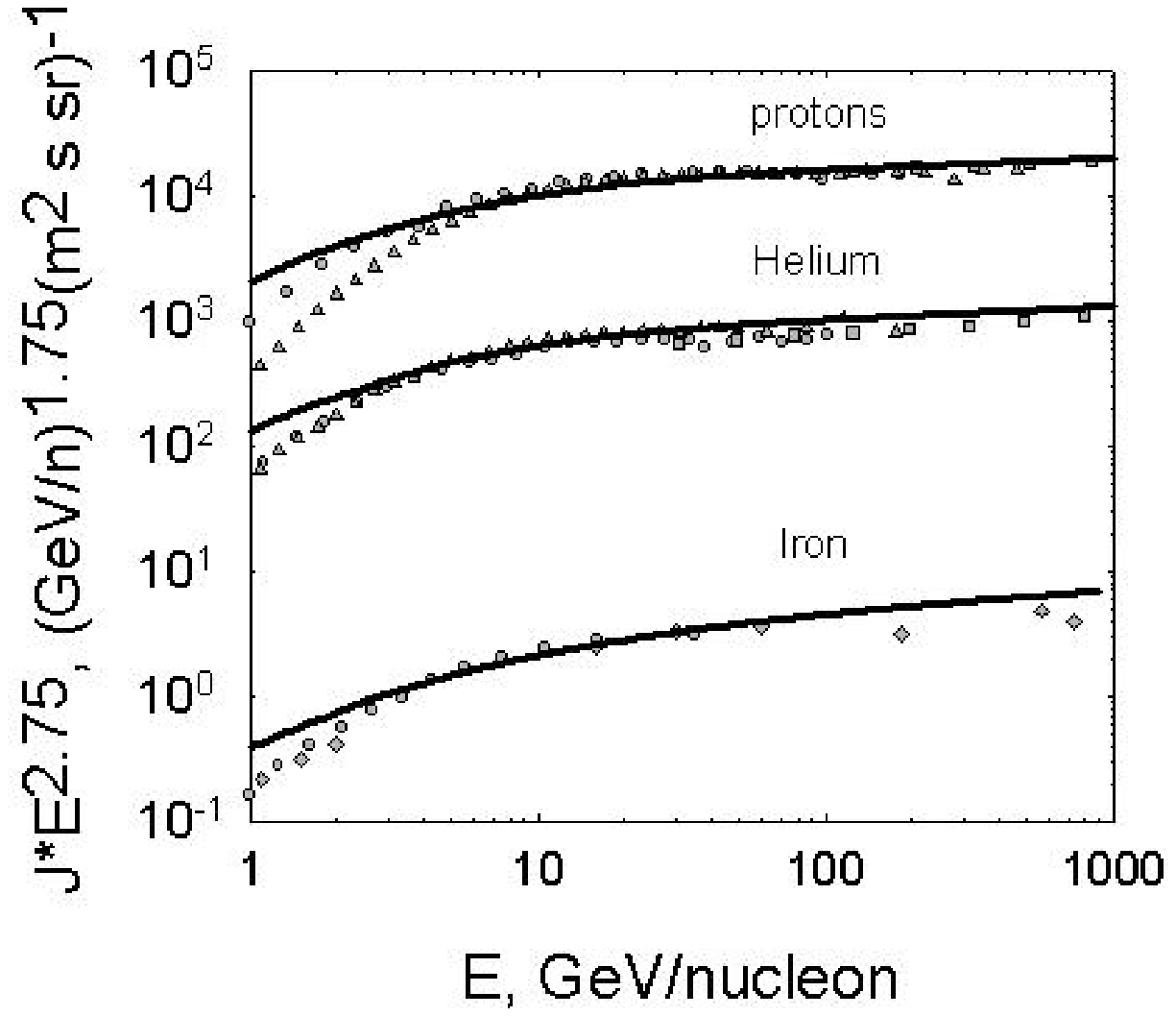}}
              \hfil
              \subfloat[]{\includegraphics[width=4.3in]{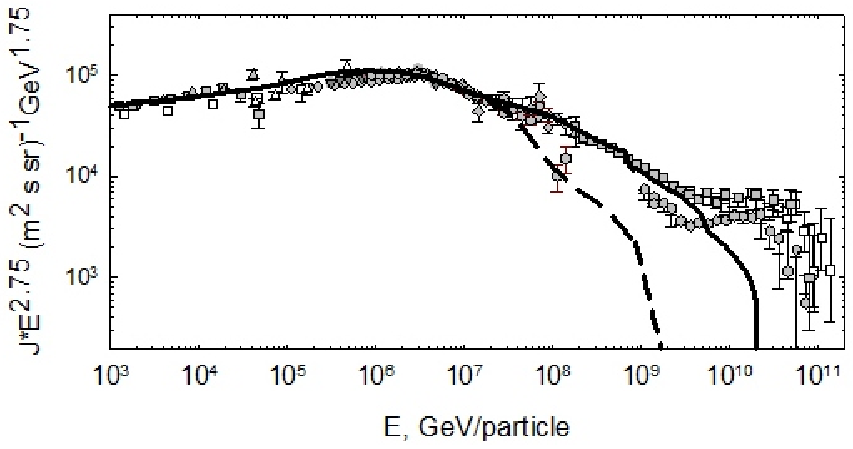}}
              }
   \caption{(a) The calculated interstellar (not corrected for solar modulation
at low energies) spectra of protons, Helium,
and Iron below energy ${10}^{3}$ GeV/n. (b) The all particle
spectrum above ${10}^{3}$ GeV; solid and dash lines are explained in the
text. See [13] for references to the
observational data shown by grey symbols.}
 \end{figure*}

The spectrum of accelerated energetic ions other than protons has
the same shape if expressed as a function of magnetic rigidity $Q(p/Z)$
with the appropriate absolute normalization determined by the injection
process at thermal energies.

Released into interstellar space from numerous supernova remnants,
the relativistic ions diffuse in galactic magnetic fields,
interact with interstellar gas, and finally escape through the
cosmic ray halo boundaries into intergalactic space, where the
density of cosmic rays is negligible. The main characteristics of
cosmic ray propagation in the Galaxy needed for the calculation of
the cosmic ray spectrum is the escape length $X_{e}$, the average
matter thickness traversed by cosmic rays before exit from the
Galaxy. Based on [14] we choose it in the form

\begin{equation}
X_{e}=\frac{11.8\left( v/c\right) }{1+{\left( p/4.9Z
\mathrm{GV}\right) }^{0.54}}\ \ g/{\mathrm{cm}}^{2}.
\end{equation}

Eq. (3) means that at high enough energies the resulting spectrum
is steeper than the source spectrum by ${0.54}$.

The results of our calculations are shown in Fig. 2a for the
interstellar spectra of protons, Helium and Iron nuclei at kinetic
energy per nucleon $1$ GeV/n$<E<{10}^{3}$ GeV/n where the charge
resolution of cosmic ray experiments is high. The combined
spectrum of all protons and ions with energies $E\geq {10}^{3}$
GeV is shown by the solid line in Fig. 2b. The source calibration
for nuclei from protons to Iron was made at one reference energy
${10}^{3}$ GeV. It was assumed that the charge composition of
accelerated particles was the same in all types of SNRs except
that the highest energy part of the spectrum produced by Type Ib/c
supernovae at $\mathrm{pc}/Z>3\cdot {10}^{5}$ GeV had no hydrogen
that ref lects the composition of highly evolved presupernovae.

The calculated spectra show remarkably good overall fit to
observations up to ultra high energies $\sim  3\cdot {10}^{9}$
GeV. To a good approximation the bending of the observed spectrum
at around the knee energy $3\cdot {10}^{6}$ GeV is reproduced
although no special efforts were made to force the theory fit the
data. The bending is due to the combined effect of the summation
over different types of SNRs and over different types of
accelerated nuclei.

The complicated chemical composition of high energy cosmic rays is
illustrated in Fig. 3 where the calculated mean logarithmic atomic
number of cosmic rays $<\ln ( A) >$ is presented. The increase of
$<\ln ( A) >$ at energies from ${10}^{5}$ GeV to ${10}^{7}$ GeV is
due to the dependence of the knee position on charge
$p_{\mathrm{knee}}\propto Z$ for each kind of ions accelerated in
Types Ia, IIP, Ib/c SNRs. Type IIb SNRs with normal composition
dominate at rigidities $p/Z>5\cdot {10}^{6}$ GV. They have a knee
at about $p_{\mathrm{knee}}/Z\approx 5\cdot {10}^{7}$ GV and
provide progressively heavier composition to the very high
energies.

\begin{figure}[htb]
\begin{center}
\includegraphics[width=2.4in]{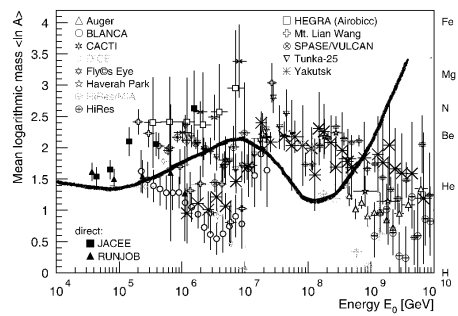}
\caption{Calculated mean logarithmic mass of cosmic rays (thick
solid line) compared to observational data based on the average
depth of shower maximum from [13].}
\end{center}
\end{figure}

The obtained cosmic ray spectrum shown by solid line in Fig. 2b is
very attractive for the explanation of cosmic ray data. However,
the use of the escape length (3) at ultra high energies is not
justified. Experimentally, the value of $X_{e} $ is determined
from the abundance of secondary nuclei in cosmic rays with good
statistics only up to about $100$ GeV/n, see [14]. If cosmic ray
transport in the Galaxy is described as diffusion, the diffusion
coefficient can be expressed through the escape length as
$D\approx v \mu  H/2X_{e}$ (here $\mu \approx 0.003
$g/${\mathrm{cm}}^{2}$ is the surface mass density of Galactic gas
disk, $H\approx 4$ kpc is the height of the Galactic cosmic-ray
halo), which gives $D\approx 1.3\cdot {10}^{28} (p c/Z
\mathrm{GeV})$ ${\mathrm{cm}}^{2} $/s. The diffusion approximation
can be applied when the diffusion mean free path $3D/v$ is less
than the size of the system $H$, which results in the condition $p
c/Z<2\cdot {10}^{7}$ GeV. The dash line in Fig. 2b shows the
result of calculations made under the assumption that cosmic ray
particles with energies $p c>2\cdot {10}^{7}Z$ GeV are lost from
the Galaxy undetected. The predicted spectrum fits observations
only up to $\sim 4\cdot {10}^{7}$ in this case.

The validity of diffusion approximation extends to higher energies
in the diffusion model with distributed reacceleration on the
interstellar turbulence where $X_{e}\propto (p/Z)^{1/3}$ at high
rigidities, see [15]. However, this scaling does not reproduce the
observed cosmic ray spectrum for the calculated source spectrum.

The physical pattern of cosmic ray propagation is different in the
models with Galactic wind. The wind model with selfconsistently
calculated cosmic-ray transport coefficients reproduces well the
data on secondary nuclei [16]. The supersonic wind is probably
terminated by the shock at $\sim 0.5$ Mpc from the Galactic disk.
The confinement of very high energy cosmic rays in the Galaxy can
be more efficient in this model as compared to the diffusion model
with a static f{}lat halo of the size $\sim 4$ kpc discussed
above. In any case, trajectory calculations in galactic magnetic
fields are needed to study cosmic ray propagation when the
diffusion approximation breaks up at ultra high energies. The
detailed consideration of this issue is beyond the scopes of the
present paper.

\section{Conclusion}

We have calculated the steady state spectrum of cosmic rays
produced by SNRs in the Galaxy. The new numerical code [5] for
modelling of particle acceleration by spherical shocks with the
back reaction of cosmic ray pressure on the shock structure was
used in the calculations. The significant magnetic field
amplification in young SNRs inferred from the observations of
thier synchrotron X-ray radiations [4], and most probably produced
by cosmic ray streaming instability, was introduced in the
calculations. It lead to the inclusion of the Alfvenic drift in
the equation for particle transport downstream of the shock. Four
different types of SNRs with relative burst rates taken from [10]
were included in the calculations. The escape length Eq.(3) from
[12] was used to describe the propagation of cosmic rays in the
Galaxy. The normalization to the observed intensity and chemical
composition of cosmic rays was made only at one energy, ${10}^{3}$
GeV.

The results are illustrated by the solid line in Fig. 2b when
Eq.(3) for $X_{e}$ is used without limitation and by the dash line
when it is limited by the applicability of the diffusion
approximation for cosmic ray propagation in the diffusion model
with a f{}lat static halo. The solid line reproduces well the
whole cosmic ray spectrum up to $\sim 3\cdot {10}^{9}$ GeV while
the dash line makes it up to $\sim 4\cdot {10}^{7}$ GeV. Further
investigations of cosmic rays propagation in galactic magnetic
fields at ultra high energies are needed to refine the predicted
shape of the spectrum produced by the Galactic SNRs. This is
important in light of the discussion about transition from the
Galactic to extragalactic component in the observed cosmic ray
spectrum [17,18].

Our results can be compared to the earlier results [19] where the
Alfvenic drift effect was not taken into account and only one type
of SNRs (the Type Ia SNRs which represent about $30$\% of all
SNRs) was considered. The absence of Alfvenic drift leads to the
very f{}lat cosmic ray source spectrum that required too strong
dependence of the escape length on rigidity $X_{e}\propto
{(p/Z)}^{-0.75}$, which is inconsistent with the available data on
secondary nuclei.

Our assumption that the composition of accelerated particles is
the same for all types of nuclei (except the very high energy part
of the spectrum produced by Type Ia SNRs) and our ignoring of the
dispersion of SNR parameters within the same type of supernovae
are probably too simplified to correspond to reality. More
comprehensive analysis with the account of earlier works [20],
[21], [22] is needed, although that requires the introduction of
new, not well known astrophysical parameters.

Another important problem is the discrete nature of supernovae in
space and time leading to fluctuations that are difficult to
account for. The number of supernovae that determines the cosmic
ray intensity is $\sim 20$ at ${10}^{8}Z$ GeV. The fluctuations of
cosmic ray intensity and anisotropy in the diffusion model were
studied in [23].

\section{Aknowledgments}

This work was supported by the RFBR grant 07-02-00028 at IZMIRAN and
by the NASA APRA grant at IPST.\\

\end{document}